# AB Levitator and Electricity Storage

**Alexander Bolonkin**
C&R, 1310 Avenue R, #F-6, Brooklyn, NY 11229, USA
(718) 339-4563, aBolonkin@juno.com, http://Bolonkin.narod.ru

## Abstract

The author researched this new idea – support of flight by any aerial vehicles at significant altitude solely by the magnetic field of the planet. It is shown that current technology allows humans to create a light propulsion (AB engine) which does not depend on air, water or ground terrain. Simultaniosly, this revolutionary thruster is a device for the storage of electricity which is extracted and is replenished (during braking) from/into the storage with 100% efficiency. The relative weight ratio of this engine is 0.01 - 0.1 (from thrust). For some types of AB engine (toroidal form) the thrust easily may be changed in any direction without turning of engine.

  The author computed many projects using different versions of offered AB engine: small device for levitation-flight of a human (including flight from Earth to Outer Space), fly VTOL car (track), big VTOL aircrat, suspended low altitude stationary satellite, powerful Space Shuttle-like booster for travel to the Moon and Mars without spending energy (spended energy is replenished in braking when ship returns from other planet to its point of origin), using AB-devices in military, in sea-going ships (submarimes), in energy industry (for example. as small storage of electric energy) and so on. The vehicles equipped with AB propulsion can take flight for days and cover distances of tens thousands of kilometers at hypersonic or extra-atmosphere space speeds.

  The work contains tens of inventions and innovations which solves problems and breaks limitations which appear in solution of these very complex revolutionary ideas.

## Introduction

   Free flight in the atmosphere as by birds was the ancient dream of people without aircraft. During 1964, the author developed his first gravity control theory [1]. The first realistic method for electrostatic levitation was offered and theoretically developed in report [2] and published in [3] Ch.15.

   However, the electrostatic levitation is possible only along special electrostatic lines. Man (car, track) cannot fly in any direction and to any place in the Earth.  Riding a vehicle with air engine (propeller or rocket) for thrust, such persons cannot even imagine or feel himself as being like a bird.

   The offered new revolutionary method does not have these defects. It allows flight in any direction in Earth's using the internal electricity storage. It does not depend on the environment of air, water, ground and does not pollute regions or the planet in any way. It can operate in vacuum, in the outer space surrounding any planet (space body) having a natural or artificial magnetic field (Earth, Mars' moon Phobos, Saturn, Magnetic Stars, White Dwarf, etc.). That can change the thrust direction without turning of the engine, levitate without expending energy. That spends energy only for lifting and to overcome air drag, but lifting energy is returning when the apparatus descends. We do not lose energy if we flight in space or at planet not having an atmosphere and return to the initial take-off place. We can free flight to Moon or Mars without loss of the summary energy from a Low Earth's Orbit if a space apparatus has a same weight. Energy spended for the acceleration of apparatus will be returned when apparatus is braking.

   This work contains conventional sections: 1. The short description of the offered AB levitator (that also may be used as friendly environmental electric car and other vihicle engine), its works; 2) Theory of innovation, and 3) Projects estimations. The first and third expositive sections are destined for non-specialist readers, the second part - for specialists.

# Brief description of new revolutionary innovention.

The offered method embraces tens inventions and innovations: magnetic devices, superconductivity devices, electricity storage, compensation of magnetic forces, use of strong matters, defence from magnetic field, concentration of magnetic field, increasing of magnetic intensity, cooling systems, in protection from space radiation, special design of magnetic and cooling devices, and many others. Some of them are briefly described below.

**1. Note from theory**. It is well-known from physics when the conductor is moved in magnetic field that has an electromagnetic force and voltage between its ends. This force is computed by equation:

$$d\bar{F} = i[d\bar{l}\,\bar{B}], \qquad (1)$$

where small vector $d\bar{F}$ is force (in Newtons) of small element $d\bar{l}$ of vector wire (in meters); $\bar{B}$ is intensity of the magnetic field, in Teslas; $i$ is electric currency, in Ampers. Square brackets [ ] note a vector production.

For straight wire the equation (1) can be written in form

$$\bar{F} = i[\bar{l}\,\bar{B}], \quad \text{or} \quad F = ilB\sin(\bar{l},\bar{B}), \quad \text{or} \quad F = ilB_n, \qquad (2)$$

where $F$ is force, N; $i$ is electric currency, A; $l$ is wire length, m; $B$ is magnetic field strength, T; $B_n$ is projection vector $\bar{B}$ on perpendicular to plate contains the vectors $\bar{l}, \bar{B}$. The direction of force may be found by left hand rule: if magnetic lines enter into the human hand's palm, the fingers show the direction of electric currency, then the pollex shows the direction of magnetic force.

The electric tension (voltage), $U$, is computed by equations

$$U = lvB_n, \qquad (3)$$

where $v$ is wire speed, m/s.

The force acts to the conductor and does not depend on wire speed. That explains why our devices can levitate without movement.

The equations (1)-(2) allow easy computation of the force and voltage of conductor in magnetic field. However, if we estimate the force in Earth's magnetic field, this force is very small. The Earth's magnetic field strength in a middle latitude equals about $B \approx 3.4\times10^{-5}$ T. If the conductor has the currency $i = 10$ A and length $l = 1$ m, the force will be only $F = 3.4\times10^{-4}$ N. That very small force is enough for moving a compass needle, but not enough for moving a larger apparatus or for flight apparatus.

**2. Design and work of AB levitator, engine and electric storage.** Author offers to overcome this difficulty by use of superconductivity, which allows a big currency up $10^5$ A/mm$^2$ and more. That means that one meter from superconductivity wire having cross-section area 3 cm$^2$ and mass 4.8 kg can create a force of about 2000 N. But superconductivity calls to mind two other main problems: the low temperature of superconductivity and a limited maximum self-magnetic field that destroys the wire's superconductibity.

These new problems we will consider later. Now, we consider the main principal difficulty: the Maxwell law says: in any closed loop electric circuit (without internal conductor) the total force (and electric intensity for constant magnetic flow) equals zero. That means the back conductor (Fig.1a) creates the same opposed force $F$ and total force will be zero. The author offers the innovation which permits avoidance of this obstacle: *to protect a back wire from Earth's magnetic field* (Fig.1b). It is known, an iron (or other good magnetic) cover adsorbs the magnetic lines (fig.1c) (see textbook "Electricity" by S.G. Kalashnikov, 1985, p. 219, fig. 167, Russian). The back wire inserted into the iron cover and insulated from it (Fig.1b).

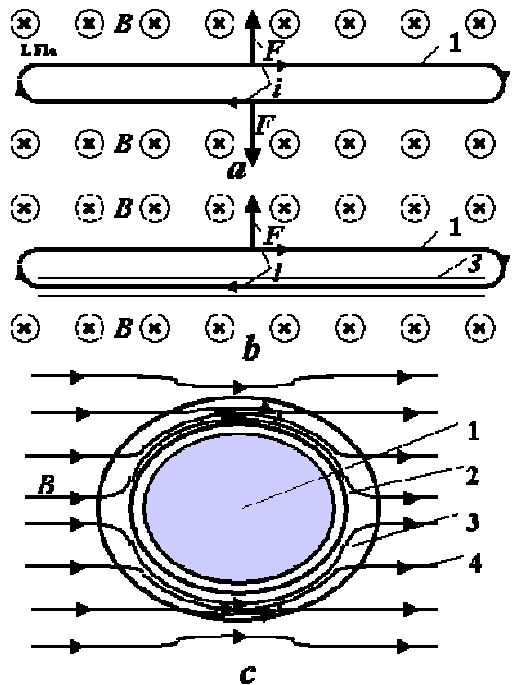

**Fig.1.** Explanation of design and work the conventional closed loop electric circuit into a magnetic field and offered AB device that creates force in the magnetic field. (*a*) Conventional electric circuit in the magnetic field. We have two opposed magnetic force *F*. (*b*) Electric circuit which has the back conductor protected from Earth's magnetic field. We have one force. (*c*) Magnetic lines flow around the back conductor. Direct exposure shield protects the back electric wire from Earth's magnetic field (cross-section of back wire 3 in "*c*").

   Notations: 1 - closed loop wire, 2 - electric insulator, 3 - protection (shield) from outer magnetic field (for example, iron, paramagnetic, ferromagnetic, etc.), 4 - lines of magnetic field, *F* - force, *i* - electric currency, *B* - magnetic field strength.

  That idea is used for AB thruster and AB levitator (Fig.2). The author offers two main forms of these devices: cylindrical form (long or shirt cylinder) and toroidal form. Naturally, each of them has advantages and disadvantages. The cylindrical form is shown in Fig. 2.

  The cylindrical AB thruster has closed loop superconductivity wires 4-6 (Fig. 2) inserted into strong composed insulator 5 (that stuff is composed from strong fibers or whiskers). That is also important **innovation** because insulator perceives the gigantic magnetic tensile forces from superconductive wires. The safety force very strong depends from this insulator. The insulator is also stores the electric energy. The amount of stored energy depends from an insulator's strength. Strong artificial fibers, whiskers, and nanotubes are preferable as the insulator (stuff). The known (for example, toroidal) superconductor electric storage has empty core. That needs in a strong heavy cover and cannot have a strong internal magnetic field. The design offered connects the straight and back wires 4, 6 in one body by the insulator and allows reaching a very strong magnetic field between wires 4 and 6 (see computation). The strong insulator also supports the superconductivity wire 4, 6 because superconductivity material has usually a small strength. The superconductive layer 11 is connected with wires 4. It is known that external magnetic field cannot penetrate the superconductive material more than $10^{-5}$ cm. The layer 11 accepts the Earth's magnetic field, passes voltage to the upper wire 4 and protects the top side of the lower back wire 6 from influence the Earth's magnetic field. The face plates 7 (or special internal cylinder made from thin ferro-magnetic (iron) material) also protects the internal side of the back wires 7 from Earth's magnetic field.

  The levitator has channels 4 where the liquid refrigerant (for example, nitrogen) circulates by a pump 9. The levitator has an innovative heat protection: multi-screens vacuum prism mirror offered by author in [3] Chs.3A, 12. The loss of refrigerant by leakage is small (see computation) and unimportant to operation. In outer space the levitator is protected from Sun and Earth radiations by the same multi-screens protector and does not need liquid refrigerant and special cooling system.

The mobile sections 10 (from thin iron (ferro-magnetic), circle and semi-circle) allows defense of part of the outer cylindrical surface from Earth's magnetic field and change (move) position and value of levitation force. The charging and discharging of the electricity storage feature (insulator between wires 4 - 6) makes inductor 8 or special outer magnetic field. We do not spend energy when the levitator levitates or moves in horizontal direction. The energy in storage is automatically spent (or replenished) when apparatus is lifting (descent), is accelerated (braking), or having movement drag (air and friction drag). That way we can free (without spending of total energy) travel in space or non-atmospheric planet if we will return at previous place and doesn't change the final weight. That capability doesn't have any known engines. The efficiency of offered electricity storage is 100% and energy saved for an unlimited time period.

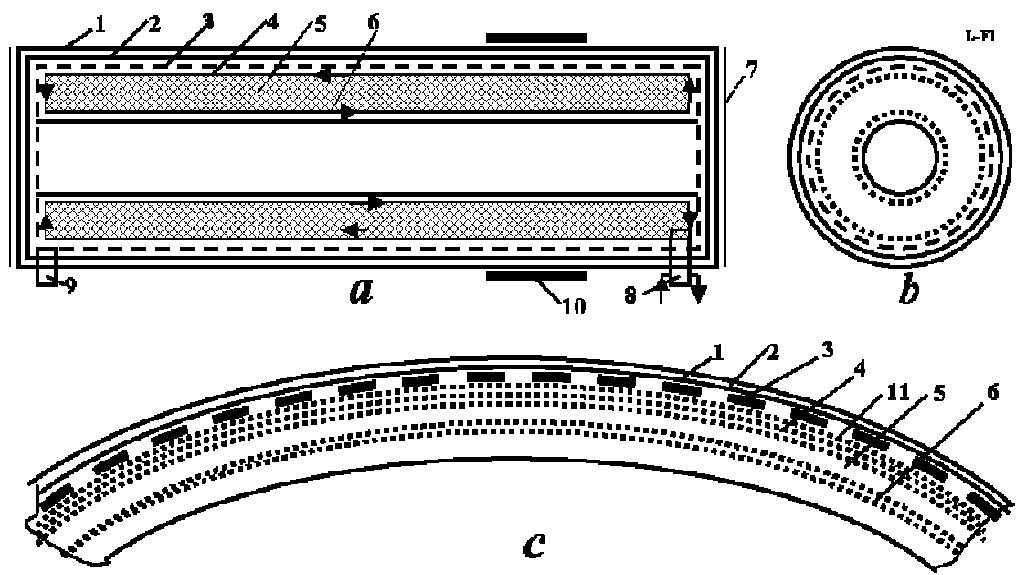

**Fig. 2**. Cylindrical superconductivity AB engine for levitator, mobile and flight apparatus and vehicles. (*a*) Side view cross-section; (*b*) Forward view; (*c*) Cross-section of forward view. Notations: 1 - cover, 2 - heat high efficiency protection (multi- high efficiency screen-mirrors); 3 - channels of the cooling system (for example by liquid nitrogen), 4 - superconductive wires inserted into the strong light insulator, 5 - strong composite insulator, 6 - back conductor, 7 - side protection from Earth's magnetic field, 8 - charge and discharge device, 9 - pump for liquid refrigerant, 10 - mobile protections of device parts from Earth's magnetic field (for control value and direction of the levitation (thrust) force), 11 - superconductive thin layer connected to wires 4. That protects the wires 6 from Earth's magnetic field.

The cylindrical levitator may be short (Fig.2) or has a thin envelope. This form is more suitable for air vehicles.

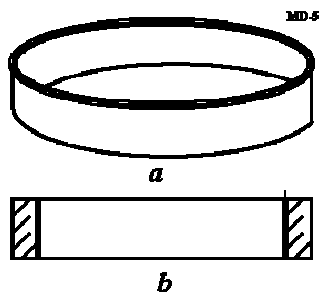

**Fig. 2**. Short cylindrical AB levitator and superconductive ring-storage of electric energy. (a) General view, (b) Cross-section of ring.

The AB levitetor can have a toroidal form (Fig. 3). This form may be better for plate flight vehicles.

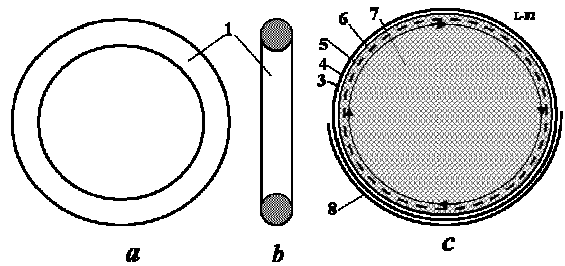

**Fig. 3**. Toroidal form of AB levitator. (a) Top view, (b) Side view, (c) Cross-section. Notations: 1 - toroidal levitator and electric storige, 3 - envelope, 4- heat protection, 5 - refrigerent, 6 - superconductive wires, 7 - strong material, 8 - protection from Earth's magnetic field and mobile control of value and direction of levitate force.

**3. Control of AB levitator**. The guidence and control of the levitating apparatis is easily accomplished. The control of the levitation force (value and direction) is presented in Fig. 4. We can slope the levitative force $F$ in any direction, to get a projection of this force to horizotal plate and move in any horizontal direction (without turning of apparatus!). Expecially that is comfortable for toroidal levitator (Fig.4h). We can also turn the apparatus around any axis. Our apparatus is neutral, but one can be stabilized quickly by the simplest gyroscopical device.

Our apparatus may be used for ground, sea ship, submarine vehicles as engine (thruster) and storage of electric energy.

The problem can be only apparent when we have **cylindrical motionless** thruster-levitator. When we move in West-East or East-West directions, we have full force $F$. When we move South-North (or back) and the main direction of an electric currency in thruster is same as the direction of the Earth's magnetic lines, the force $F$ may be closed to zero (see eq. (1)). More exactly, a butt-end of the levitate cylinder crosses the magnetic lines and produce the force, but usually the butt-end area is significantly less than the side area of the cylinder and this force is small. Only toroidal levitator has same $l$ in any direction. This defect may be corrected by many means. For example, the AB engine is turning and have West-East direction for any speed direction of vehicle; the vehicle has ferro-magnetic antennas (which change the local direction of the Earth's magnetic lines, Fig.4h)., the vehicle has zig-zag way, and so on. The toroidal thruster (Fig. 3) does not have this problem.

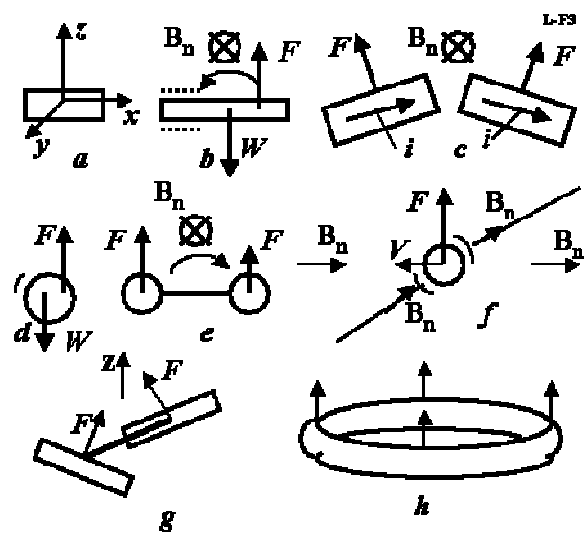

**Fig.4**. Control of value and direction of the levitation force. (*a*) Axis of cylindrical levitator. (*b*) The closing of part cylinder from Earth's magnetic fields (left end) moves the levitate force $F$ from the center of gravity and creates the moment around axis $y$. (*c*) The slope of cylinder bends the force $F$ and creates the horizontal force. (*d*) The closing of part cylinder from Earth's magnetic field moves the $F$ from the center of gravity and creates the moment around axis $x$. (*e*) The different forces $F$ of two connected

cylinder create the moment around axis *x*. (*f*) Ferromagnetic antenns returns the Earth's magnetic field near levitator and produce force when the levitator moves along the lines of Earth's magnetic field. (*g*) The opposed slope of two connected cylinders create the moment around axis *z*. (*h*) Toroidal levitator. The closing of different parts of toroid create the different forces in these parts. That allows creating the slope of apparatus and horizontal force in any direction (include the value of the force).

**4. Ferro-magnetic concentrator of outside magnetic lines**. This innovation may be very useful. Ferro-magnetic matter collects the magnetic lines (magnetic field, magnetic flow). The electromagnetic induction is

$$B = \mu\mu_0 H ,  \qquad (4)$$

where *B* is magnetic induction in T, $\mu$ is magnetic permeability (in vacuum $\mu = 1$), $\mu_0 = 4\pi \times 10^{-7}$ is magnetic constant in N/A$^2$, *H* is magnetic induction in A/m.

Magnetic induction at the Earth's magnetic equator is 27.1 A/m. The maximum magnetic permeability are: for iron $\mu = 6100$. for the steel Э 310 $\mu = 36,000$ at $H = 9.6$ A/m (maximum $B = 1.75$ T); for permalloy (having 78.5 Ni) $\mu = 100,000$ at $H = 2$ A/m, ($B_{max} = 0.8$). For iron (having 4.3% Si) $B = 0.45$ T at $H = 40$ A/m.

In case of a correct design the permeability can increase magnetic field by hundreds or thousands of times. It is used in ferro-magnetic antennas in small radio receivers. That also may be used in AB levitator, thruster, and engine for increasing the propulsion force. It is very important that the ferro-magnetic changes the direction of the magnetic lines near itself. The radius of collection (when magnetic wire is parallel to magnetic lines) equals approximately a length of the wire. The typical levitator with magnetic booster is shown in Fig. 5. However, ferromagnetic method still needs more thorough, detailed investigation.

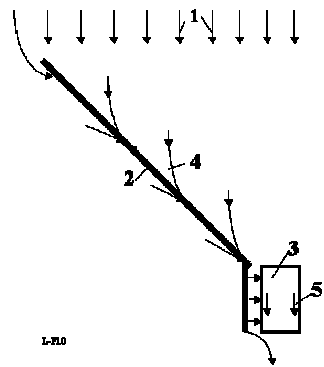

**Fig. 5**. Ferro-magnetic collector of outside magnetic field. Notation: 1 - outside magnetic field, 2 – ferro-magnetic circuit, 3 - levitator, 4 - magnetic lines, 5 - electric currency.

**5. Cooling system.** Fig. 6 shows some innovative methods of protection for the AB levitator in outer space (in vacuum). The simplest protection is a super reflective mirror (Fig. 6a) offered by author in 1988 (see [3] Chs. 12, 3A). The superreflectivity screen-mirror gives deep cooling. The usual high reflectivity screen-mirror gives enough cooling (Fig. 6b). The usual multi-screens protection also gives enough cooling (Fig. 6c)(see the computation in Theoretical Section). The Earth's atmospheric levitator needs liquified nitrogen (or cheaper liquid air, oxygen) cooling system, but by using the super reflectivity multi-screen protection we have a very small heat flow and need very small amount of liquid nitrogen.

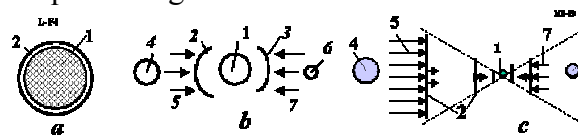

**Fig. 6.** Methods of cooling (protection from Sun radiation) the superconductivity levitator (driver, thruster) in outer space. (*a*) Protection the levitator (engine) by the high super refletivity mirror. (*b*) Protection by high reflectivity screen (mirror) from Sun and Earth's radiation. (*c*) Protection by usual multi-screens. Notations: 1 - superconductive wires (engine); 2 - heat protector (super reflectivity mirror in

Fig.6a and a usual mirror in Fig. 6c); 2, 3 - high reflectivity mirrors (Fig. 6b); 4 - Sun; 5 -Sun radiation, 6 - Earth (planet); 7 - Earth's radiation.

**6. Magnetic safety for Human life**. The intensity of magnetic field used by the levitator is very high up 60 - 180 T. Is this dangerous for human health? My answer: if we have the correct design, the offered levitator contains **all** magnetic field inside the levitator (Fig. 7). Outside of the levitator the magnetic field equals zero!

When the levitator design is correct, the magnetic field 6 from the internal (lower) circuit equals and opposed the magnetic field 5 from the outer (top) circuit is outside of the levitator. The magnetic field is located only between the lower and top circuits. Our design of superconductivity electric storage is different from common superconductivity ground storages. The first innovation is filling the internal core by a strong matter (stuff) which can keep the high tensile stress, the second innovation - we have only ONE turn (coil) of wire (the current superconductivity electric storages have a lot of turns and, as result, they are outward of the magnetic field).

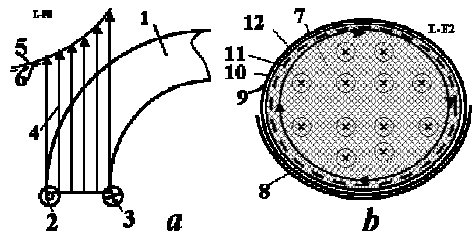

**Fig. 7.** The magnetic field of the levitator. (*a*) Cylindrical levitator, (*b*) Cross section of the toroidal levitator. Notation: 1 - is cross section of cylindrical levitator; 2 - wire of top part of coil, 3 - wire of lower part of coil; 4 - magnetic lines; 5 - magnetic intensity fron lower part of coil; 6 - magnetic intensity from top part of coil (with minus); 7 - internal core from material (dielectric) having a high tensile stress; 8 - screen from outside (Earth) magnetic field, 9 - envelope; 10 - heat protection; 11 - cooling system; 12 - superconductivity electric wire and electric currency.

**6. Artificial Magnetic field.** The possibility of levitation can be increased thousands times by the artificial magnetic field. If the magnetic field has enough intensity, the people, car can fly without non-superconductivity levitators, the ground car can receive electric energy from variable magnetic field (it is a solution of a problem of an invironment and oil), the city resident can receive the electricity without wires, orbiting satellites and spaceships can receive the thrust and energy when they flyby over this region during their flights. That may be useful for big city having an dense street traffic. The computation of this case is in the Theoretical section and estimations in the Project section.

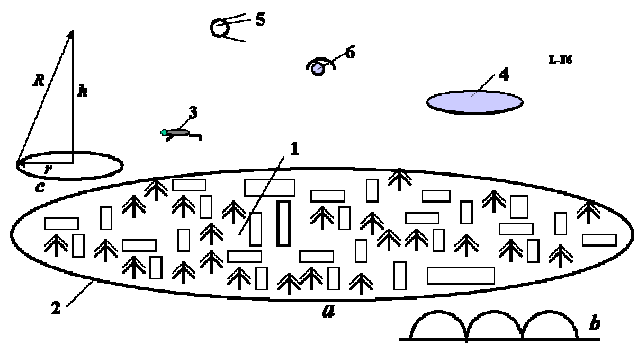

**Fig. 8.** Artificial magnetic field. (*a*) City having the superconductivity electric ring. (*b*) Variation intensity of the magnetic field. Notation: 1 - city, 2 - superconductivity electric ring, 3 - fligth man, 4 - levitator in the form of a flight plate (UFO), 5 - Earth's satellite, 6 - levitative illumination lamp. $R = \rho$ in Eq. (16).

# 3. Theory and Computation

**1. Computation (estimation) of the levitation force, storige energy, and weight of thruster.**
For estimation, the following equations of magnetodynamics are used:

$$F = ilB_n, \quad B = \mu_0 H, \quad H = i/2\pi R, \tag{5}$$

where all magnitudes are noted in the equations (1)-(4). The new magnitude $R$ is the internal radius of cylindrical tube or toroid, m.

The important values presented the AB levitator also can be received from magnitodynamics. They are

$$w_v = \frac{B^2}{2\mu_0}, \quad w_m = \frac{w_v}{\gamma_S}, \quad v_{max} = \sqrt{2w_m} = \frac{B}{\mu_0 \gamma_S}, \quad H_{max} = \frac{w_m}{g}, \tag{6}$$

where $w_v$ is volume density of energy, J/m³, that is also internal pressure into energy storage in N/m²; $w_m$ is mass density of energy releted to wire mass, J/kg; $\gamma_s$ is density of stuff, kg/m³; $v_{max}$ is impulse of levitator stuff, m/s; $H_{max}$ is maximum altitude which the stuff can self-lift at Earth, m. These values show the energetic properties of syperconductivity stuff. Computation of them are presented in Figs. (9)-(11).

The mass of the levitation engine is sum of stuff, wire, cover and cooling and control systems. If engine is intended for great acceleration and does not need large storage of energy, the main mass is wire. If device is designed as big storage of electric energy, the main mass is stuff (80 - 95%). If device operates in outer space, we do not need a cooling system; in the atmosphere the cooling system increases the mass of the engine by 10 - 30%.

For engine having large storage energy, the magnitudes of Eq. (6) is better computed for $\gamma_s$ - density of stuff. They may be also computed for $\gamma_w$ is specific mass of wire (coil of levitator). The value of Eq. (6) may be computed for $\gamma$ - density of engine. In this case $v_{max}$ and $H_{max}$ is maximum speed and maximum altitude which can reach engine without useful load.

Below are computations of Eq, (6).

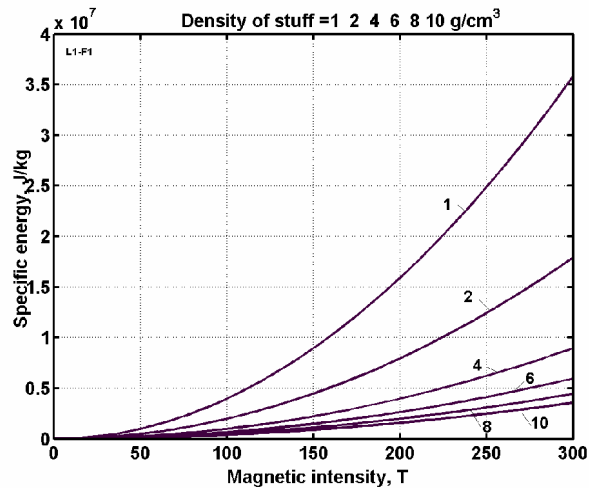

**Fig. 9.** Specific mass energy of superconductivity AB levitation engine. These result may be applied to stuff, wire, and engine.

The AB storage energy is similar to the heating value of a conventional fuel or the specific energy of a rocket fuel. Automotive gasoline has heating value $40 \times 10^6$ J/kg, but internal combustion engine has efficiency coefficient about 0.3. That way we use only $12 \times 10^6$ J/kg. More over, the combustion engine requires oxygen (air) and contaminates the Earth's atmosphere. The liquid rocket fuel (fuel + oxidizer) has about $12 \times 10^6$ J/kg, the solid-rocket fuel has about $8 \times 10^6$ J/kg. As you see (Fig. 9), the superconductive storage is the same with the rocket fuel. but it is sufficiently more for B > 250 T. Our advantage is 100% efficiency, absence of air pollution and multi-using, because we can recharge the storage device many times.That has also good prospect because then we can have greater safety maximum *B* and more strong stuff of energy storage in the future (nanotubes).

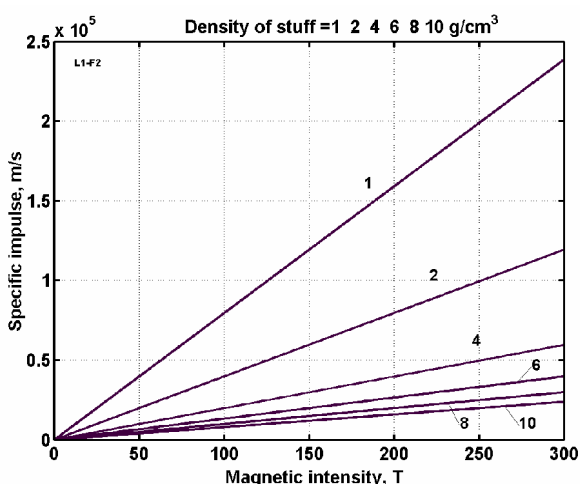

**Fig. 10.** Specific impulse of supercondutivity engine. These results may be applied to wire, stuff and engine.

The specific impulse of current rocket fuel is about 2000 ÷ 2500 m/s for solid fuel, 3000÷3200 m/s for liquid fuel, and up 4000 m/s for hydrogen fuel. In our case, impulse is about 50,000 ÷ 100,000 m/s, in 20 ÷ 40 times more (Fig. 10). Some electric propulsion has high impulse (10,000÷30,000 m/s), but they are technically complex, need large electric energy, and produce only a small thrust. Our propulsion contains energy in its electric storage, can have a big thrust and can be charged distantly from the Earth by artificial magnetic field.

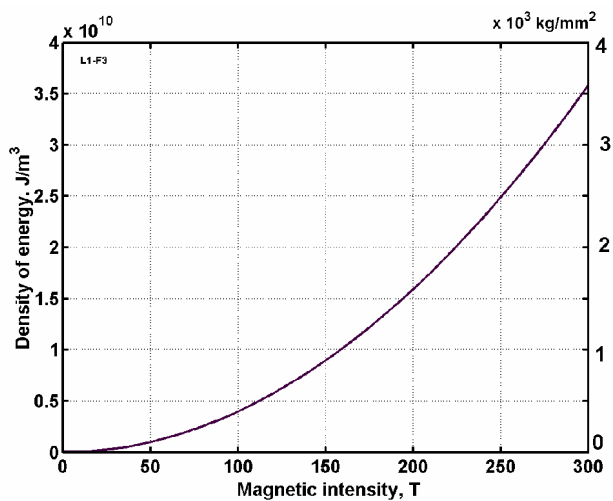

**Fig. 11.** Volume energy of superconductivity storage. The right scale shows the internal pressure into energy storage.

Fig. 11 shows the volume density of energy and magnetic pressure into magnetic storage. Before $B =$ 100 T we can use the conventional strong fibers as internal stuff of the superconductive storage (see Table 2 below), in interval $B = 100÷250$ T we must use whiskers, and over $B \geq 250$ T we need in nanotubes.

**2. Data.** *Magnetic field*. Magnetic induction at the Earth's magnetic equator is 27.1 A/m ($B = 3.4 \times 10^{-5}$ T), at the magnetic pole $H = 52.5$ A/m. In some regions (for example, near Kurs, Russia) $H \approx 100$ A/m. Magnetic intensity at very high altitude is presented in Fig. 12, [5], p. 133. We will use $B = 3.4 \times 10^{-5}$ T in our projects up to the altitude 500 km. The closed magnetic field has the small Martian satellite Phobos. The Sun, Mars and Jupiter have magnetic fields and magnetic stars have powerful fields too. The White Dwarf star has B ≈ 80,000 T.

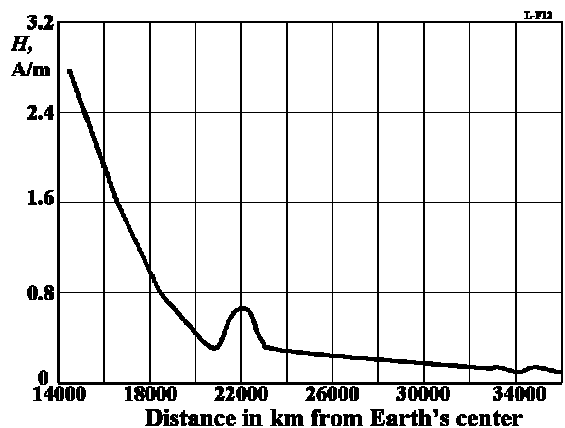

**Fig. 12.** Earth's magnetic intensity far from the Earth's center.

There are hundreds of new superconductivity matters (type2) having critical temperature 70 ÷ 120 K and more.
*Some of the superconductable material* are presented in Table 1 (2001). The widely used $YBa_2Cu_3O_7$ has mass density 7 g/cm$^3$.

**Table 1**. Transition temperature $T_c$ and upper critical field $H_{c2}(0)$ of some examined superconductors [4], p. 752.

| Crystal | $T_c$ (K) | $H_{c2}$ (T) |
|---|---|---|
| La $_{2-x}$Sr$_x$CuO$_4$ | 38 | ≥80 |
| YBa$_2$Cu$_3$O$_7$ | 92 | ≥150 |
| Bi$_2$Sr$_2$Ca$_2$Cu$_3$O$_{10}$ | 110 | ≥250 |
| TlBa$_2$Ca$_2$Cu$_3$O$_9$ | 110 | ≥100 |
| Tl$_2$Ba$_2$Ca$_2$Cu$_3$O$_{10}$ | 125 | ≥150 |
| HgBa$_2$Ca$_2$Cu$_3$O$_8$ | 133 | ≥150 |

The last decisions are: Critical temperature is 176 K, up 183 K. Hanotube has critical temperature 12 - 15 K,
Some organic matters has temperature up 15 K. Polypropylene, for example, is normally an insulator. In 1985, however, researchers at the Russian Academy of Sciences discovered that as an oxidized thin-film, polypropylene can have a conductivity $10^5$ to $10^6$ higher than the best refined metals.
Boiled temperature of liquid nitrogen is 77.3 K, air 81 K, oxygen 90.2 K, hydrogen 20.4 K, helium 4.2 K [6].
Unfortutately, most superconductive material is not strong and needs a strong covering.

*Material strong*. Let us consider the following experimental and industrial fibers, whiskers, and nanotubes:
1. Experimental nanotubes CNT (carbon nanotubes) have a tensile strength of 200 Giga-Pascals (20,000 kg/mm$^2$). Theoretical limit of nanotubes is 30,000 kg/mm$^2$.
2. Young's modulus is over 1 Tera Pascal, specific density $\gamma=1800$ kg/m$^3$ (1.8 g/cc) (year 2000). For safety factor $n = 2.4$, $\sigma = 8300$ kg/mm$^2$ = $8.3\times10^{10}$ N/m$^2$, $\gamma=1800$ kg/m$^3$, $(\sigma/\gamma)=46\times10^6$. The SWNTs nanotubes have a density of 0.8 g/cm$^3$, and MWNTs have a density of 1.8 g/cm$^3$ (average 1.34 g/cm$^3$) Unfortunately, the nanotubes are very expensive at the present time (1994).
3. For whiskers $C_D$ $\sigma = 8000$ kg/mm$^2$, $\gamma = 3500$ kg/m$^3$ (1989) [3, p. 33]. Cost about $400/kg (2001).

4. For industrial fibers $\sigma = 500 - 600$ kg/mm$^2$, $\gamma = 1800$ kg/m$^3$, $\sigma/\gamma = 2{,}78 \times 10^6$. Cost about 2 - 5 $/kg (2003).

Figures for some other experimental whiskers and industrial fibers are given in Table 2.

Table 2. Tensile strength and density of whiskers and fibers

| Material Whiskers | Tensile strength kg/mm$^2$ | Density g/cm3 | Fibers | Tensile strength kg/mm$^2$ | Density g/cm$^3$ |
|---|---|---|---|---|---|
| AlB$_{12}$ | 2650 | 2.6 | QC-8805 | 620 | 1.95 |
| B | 2500 | 2.3 | TM9 | 600 | 1.79 |
| B$_4$C | 2800 | 2.5 | Thorael | 565 | 1.81 |
| TiB$_2$ | 3370 | 4.5 | Alien 1 | 580 | 1.56 |
| SiC | 2100-4140 | 3.22 | Alien 2 | 300 | 0.97 |
| Al oxide | 2800-4200 | 3.96 | Kevlar | 362 | 1.44 |

See Reference [3] p. 33.

**3. Computation (Estimation) method**. For estimation AB levitator the author recommends the following method based on Eq. (4) and on usial equations for volum, mass area, etc:

1) Take the need payload (or mass needs in movement) $m_p$, safety magnetic intensity $B$, and dencity storge stuff and wire $\gamma_s$, $\gamma_w$ (You can use the Tables 1, 2). You find by Eq. (5) (or by Figs. 9-11) the magnitudes $w_v$, $w_m$, $v_{max}$, $H_{max}$. Check up the tensile stress of stuff from the internal magnetic pressure $p = w_v$ in N/m$^2$ or $p = 10^{-7} w_v$ in kg/mm$^2$ (Fig. 9).

2) Take the need flight data: altitude $H < H_{max}$, speed $V < v_{max}$, range $L$ ($H$ in m, $V$ in m/s, $L$ in m). The air drag $D$ (in N) of your vehicle, the full mass $m_x$ (in kg), or stuff mass $m_s$ (in kg) of your vehicle may be estimated by equations:

$$D = C_x \frac{\rho V^2}{2} S, \quad m_x = \frac{m_s w_m - DL - E_{in}}{0.5 V^2 + gH}, \quad m_s = k \frac{m_p(0.5 V^2 + gH) + DL + E_{in}}{w_m - 0.5 V^2 - gH}, \quad (7)$$

where $C_x \approx 0.02 \div 0.7$ is coefficient of air drag; $\rho = 1.225$ kg/m$^3$ is air density at altitude $H \approx 0$; $S$ is typical area of vehicle, m$^2$, for example, cross-section or wing area; $k \approx 1 \div 2.1$ reserve coefficient for lift force; $E_{in}$ expenditure the energy for internal needs (for example, light), in J/kg; $g = 9.81$ m/s$^2$; $m_p$ is useful mass of the vehicle, kg.

Equation (7) are recived from an energy balance of the levitation vehicle for $k = 1$.

$$m_x(0.5 V^2 + gH) + DL + E_{in} = w_m m_s, \quad m_x = m_p + m_s . \quad (8)$$

Look your attention in the second Eq. (7): if we do not move our apparatus, the mass it may be infinity. That means we can suspended our device (for example translator) in given point ($V = 0$) and does not spend energy its support (exsept stabilization and cooling).

3) Find the minimal (internal) radius $R_m$ of levitator (AB engine) and minimal currency $i_m$ for lift force $F = gm_x$, $l = 2h$:

$$R_m = \frac{\mu_0 F}{4\pi hBB_n}, \quad i_m = \frac{F}{2hB_n}, \quad (9)$$

where $h$ is the length (height) of levitator cylinder or small diameter of toroid, m.

4) Find volume $v$, mass $m_s$ (weight), and thickness of stuff $\delta$ of the cylindrical storage stuff:

$$\delta = \frac{m_S}{2\pi Rh\gamma_S}, \quad \text{or} \quad E = (m_x - m_p)w_m, \quad v = \frac{E}{m_v}, \quad m_S = v\gamma_s, \quad \delta = \frac{v}{2\pi Rh}, \quad r = \frac{1}{\pi}\sqrt{\frac{v}{2R}}, \quad (10)$$

where $E$ is energy into electric storage, J; $R > R_m$ is average radius of cylinder or toroid, m; $\gamma_s$ is density of storage stuff, kg/m$^3$, $r$ is small radius of toroid, m.

5) Find the cross-section $s$ and mass $m_w$ of superconductivity wire:

$$s = i / j, \quad m_w \approx 2(h + \delta)s\gamma_w, \quad (11)$$

$j \approx 10^5$ A/mm$^2$ = $10^{11}$ A/m$^2$ is dencity of the electric currency in superconductivity wire; $\gamma_w$ is density of superconductivity wire, kg/m$^3$ (2000 ÷ 8000 kg/m$^3$).

6) Total weight of levitation engine is sum of mass stuff plus mass wire $m = m_s + m_w$. It must be increased in 10 ÷ 20% for cooling system and cover.

*Example of computation*. **Levitation of a flying human.**
We want to design a levitation belt to lift a living human weighing 82 kg (Fig. 3).
1) Take the magnetic field into AB levitator $B = 60$ T; wire density $\gamma_w = 8000$ kg/m$^3$; For $B = 60$ T the magnetic internal pressure is $p = 10^{-7} w_v = 140$ kg/mm$^2$. As the stuff may be taken the fiber (Table 2). the stuff density $\gamma_s = 1800$ kg/m$^3$, $\sigma = 600$ kg/mm$^2$. The computation gives (Eq. (5) or Figs. (9)-(11) ): $w_v = 1.4 \times 10^9$ J/m$^3$, $w_m = w_v/\gamma_s = 7.8 \times 10^5$ J/kg .
2) Take the flight speed $V = 15$ m/s = 54 km/s, range $L = 100$ km =100.000 m, altitude $H = 100$ m, $E_{in} = 0$, $k = 2$, $h = 0.1$ m. Compute the air drag: $D = 1.4$ N and stuff mass $m_s = 0.36$ kg (Eq. (7)).
3) Compute minimal radius and currency in belt: $R_m = 0.2$ m, $i = 1.8 \times 10^8$ A (Eq. (7)).
4) Compute volume, weight, and thickness of AB levitate belt (Eq. (8)): $\delta = 1.6$ mm .
5) Compute cross-section and weight of superconductive wire (Eq. (9)) for density of electric currency $j = 10^5$ A/mm$^2$: $s = 1.18 \times 10^{-3}$ m$^2$ , $m_w = 1.89$ kg.

The sum mass of stuff and wire is about $0.36 + 1.89$ kg $\approx 2.25$ kg. This levitator mass we must increase about in 0.5 - 1 kg for cover, control and cooling system. The total mass of levitator is about 3 kg.

The internal diameter is about 39 cm, outer diameter is about 42 cm, height of cylinder (belt) is about 12 cm (Fig.3). Remainder the total lift force (together with man) is 82 kg.

These data are not optimal but acceptable for a market individual small vehicle (belt). The apparatus is based on current technology and can be made at present time (the many offered innovations and inventions must be used).

**5) Computation of AB Levitation Launcher**. For $B = 140$ T specific impulse of AB engine equals the specific impulse of a liquid rocket engine. However one quickly decreases (as in the second power) when $B$ increases. The tensile stress of the stuff also rapidly increases. For $B = 140$ T it equals 780 kg/mm$^2$. That is acceptable for whiskers having maximum $\sigma$ up $4000 \div 8000$ kg/mm$^2$. The very high $B > 200$ T is very efficiency as electric storage but request nanotubes as the stuff.

The levitation launch has a many advantages, but in difference of rockets that does not spend a fuel mass. The weight of AB-engine is same in beginning and ending acceleration. As a result, the AB engine for high acceleration apparatus needs many stages of multi-launcher.

I recommend the following order of the estimation. Initially we do not include the wire mass, cover and the cooling system. Their mass is only 10 - 15% of the stuff mass. We take into account when we take the increased final useful weight of apparatus in 10 - 15%. They are included into in increased mass of stuff.

The initial data are: the mass $m_p$, final speed $V$, and final altitude $H$ of apparatus (satellite, spaceship, or interplanetary probe). We take the equals distribute speed and altitudes between stages of the levitation accelerator.

In these conditions the mass of every engine stages may be estimated by equations:

$$m_n = a^n m_p, \quad \text{where} \quad a(N) = \frac{w_m}{w_m - 0.5(V/N)^2 - gH/N}, \quad n = 1,2,3,...,N, \quad (12)$$

where $N$ is number of stage ($n$ is numbering from payload). This ratio is received from balance of energy.

The mass of stuff included mass wire, cover and cooling system in every stage may be estimated as below

$$m_s(n) = m_n / a(N) \quad \text{or} \quad m_s(n) = m_p / a^n, \quad n = 1,2,3,...,N. \quad (13)$$

The other values ($R_m$, $i_m$, $\delta$, $s$) are computed by equations above.

The example of computation the AB launcher for launching of spaceship $m = 20$ tons to Moon and Mars is in Projects section.

**6) Artificial of magnetic field.** The capability of levitation apparatus may be increased thousand times if we create the outside strong artificial magnetic field. In powerful artificial magnetic field we can achieve sustained flight without superconductivity devices. This field is useful in big city having heavy traffic congestion. The artificial magnetic field may be created by superconductivity ring of a large diameter. This field may be also useful for spaceship. When they flyby into artificial magnetic field the force increases hundreds of times.

The data of the artificial magnetic field and flight apparatus may be computed by equations:

$$F = ilB = jlsB = jvB, \quad v = ls = m/\gamma, \quad j = i/s, \quad r = \rho l/s, \quad E = i^2 r, \qquad (14)$$

where $F$ is lift force, N; $i$ is electric currency, A; $l$ is length of AB engine wire into open magnetic field, m; $B$ is intensity of the artificial magnetic field, T; $j$ is density of electric currency of AB engine, A/m$^2$; $v$ is wire volume of engine, m$^3$; $\gamma$ is wire mass density, kg/m$^3$; $s$ is cross-section of wire, m$^2$; $r$ is wire electric resistance, $\Omega$; $\rho$ is specific resistance of wire, $\Omega\cdot$cm; $E$ is loess of energy into conventional (non-superconductivity) AB engine, J; $m$ is wire mass, kg.

From Eq. (14) we can receive the following ratios for estimation of a data of the levitation vehicles (do not having the superconductive wires):

$$\frac{F_{kg}}{m} = \frac{jB}{2g\gamma}, \quad \frac{E}{m} = \frac{2\rho j^2}{\gamma}, \qquad (15)$$

where $F_{kg}$ is lift force in kg. The number "2" appears because the coil has no active back wire. Aluminum wire has $\rho = 2.8 \times 10^{-8}$ $\Omega\cdot$m (at room temperature) and mass density $\gamma = 2800$ kg/m$^3$. The maximum of a currency dencity for non-cooling wire is about $j = 10^7$ A/m$^2$ (10 A/mm$^2$). It is obvious, the first ratio in (14) must be over 1 for the levitation apparatus.

The intensity $H$ of the artificial magnetic field at altitude (Fig. 8c) may be computed by formula

$$H = \frac{iS}{2\pi \rho_h^3}, \qquad (16)$$

where $S$ is area of the ground closed loop ring (enveloped by electric currency), m$^2$; the distance $\rho_h$ equals approximately altitude for high height (Fig.8c, $R = \rho_h$).

Weakness of artificial magnetic field is vertical direction of magnetic lines near ground surface. The magnetic force has horizontal direction. The levitation apparatus needs magnetic antennas (Fig. 5) for lift force creation (or other method for changing the direction of the magnetic lines).

The artificial magnetic field and apparatus lift force may be created by permanent magnets.

The example of computation the levitation device for the flying individual human into artificial magnetic field are presented in section "Projects".

**7) Computation of the cooling system**. The following equations allow direct computing of the proposed cooling systems.

1) Equation of heat balance of a body in vacuum

$$\zeta q s_1 = C_s \varepsilon_a \left(\frac{T}{100}\right)^4 s_2, \qquad (17)$$

where $\zeta = 1 - \xi$ is absorption coefficient of outer radiation, $\xi$ is reflection coefficient; $q$ is heat flow, W/m$^2$ (from Sun at Earth's orbit $q = 1400$ W/m$^2$); $s_1$ is area under outer radiation, m$^2$; $C_s = 5.67$ W/m$^2$K is heat coefficient; $\varepsilon_a \approx 0.02 \div 0.98$ is blackness coefficient; $T$ is temperature, K; $s_2$ is area of body or screen, m$^2$.

2) Heat flow between two parallel screens

$$q = C_a \left[\left(\frac{T_1}{100}\right)^4 - \left(\frac{T_2}{100}\right)^4\right], \quad C_a = \varepsilon_a C_s, \quad \varepsilon_a = \frac{1}{1/\varepsilon_1 + 1/\varepsilon_2 - 1}, \qquad (18)$$

where the lower index $_{1,2}$ shows (at $T$ and $\varepsilon$) the number of screens. Every coventional screen decrease temperature aproximately in two times.

3) When we use the conventional heat protection the heat flow is computed by equations

$$q = k(T_1 - T_2), \quad k = \frac{\lambda}{\delta}, \qquad (19)$$

where $k$ is heat transmission coefficient, W/m$^2$K; $\lambda$ - heat conductivity coefficient, W/m·K. For air $\lambda$ = 0.0244, for glass-wool $\lambda$ = 0.037; $\delta$ - thickness of heat protection, m.

**Table 3**. Boiling temperature and a heat of varoporation of some relevant liquids [6]. p.68.

| Liquid | Boilng temperature, K | Heat varoparation, kJ/kg |
|---|---|---|
| Hydrogen | 20.4 | 472 |
| Nitrogen | 77.3 | 197.5 |
| Air | 81 | 217 |
| Oxygen | 90.2 | 213.7 |
| Carbonic acid | 194.7 | 375 |

These data are enough for computation of the cooling systems.

Using the correct design of multi-screens, high reflectivity screen, and vacuum between screens we can get a very small heat flow and a very small expenditure for refrigerant (some grams per day in Earth). In outer space the protected body can have low temperature without special liquid cooling system (Fig.6),.

For example, the space body (Fig. 6a) with innovative prism reflector [3] Ch. 3A ($\rho = 10^{-6}$, $\varepsilon_a$ = 0.9) will have temperature 13 K in outer space. The protection Fig.6b gives more low temperature. The usual multi-screen protection of Fig. 6c gives the temperature: the first screen - 160 K, the second - 75 K, the third - 35 K, the fourth - 16 K.

## 4. Projects

**1. Flying (levitation) human** was computed above in Theoretical section..

**2. Flying car** (see plan and equations above):
1. Take the data: magnetic intencity $B$ = 60 T, $\gamma_s$ = 1800 kg/m$^3$, $\gamma_w$ = 8000 kg/m$^3$, form is **two** cylinders $h$ = 1.5 m, $B_n$ = 3.4×10$^{-5}$ T.
2. Computation: $w_v$ = 1.4×10$^9$ J/m$^3$, $P$ = 140 kg/mm$^2$, $w_m = w_v/\gamma_s$ = 7.8×10$^5$ J/kg .
3. For $V$ = 20 m/s = 72 km/h, $H$ = 1000 m, $L$ = 1000 km = 10$^6$ m, $E_{in}$ = 7.2×10$^5$ W, $C_x$ = 0.08, $S$ = 1.5 m$^2$, we receive $D \approx$ 30 N, $m_s$ = 52 kg for one cylinder.
4. Minimal cylinder radius and currency (for lift force one cylinder $F$ = 5500 N): $R_m$ = 0.172 m, $i$ = 5.16×10$^7$ A.
5. Thickness of stuff: $\delta$ = 17.8×10$^{-3}$ ≈ 18 mm.
6. Cross-section and mass of wire (one tube) for currency density $j$ = 10$^5$ A/mm$^2$: $s$ = 5.16 cm$^2$, $m_w$ = 12.4 kg.
7. Total mass of levitation engine (two tubes): $m$ = 1.1×2(52+12.4) ≈ 142 kg.

The levitation engine may be designed as two tubes which join to any CURRENT cars. The internal combustion engine, transmission, fuel tank may be removed.

The offered AB-engine not only saves the planetary invironment, releases a country from oil dependence, one spends energy sometimes less than when using a liqued fuel. The drag of usual car endures the friction of its wheels on the ground and some air drag. For friction coefficient 0.1 the friction drag for car mass 1000 kg is about 1000 N plus air drag 30 - 100 N. The flying car does not have wheel friction. That means AB car spends an energy for moving that is 20 times less than a conventional car. Make corrections that internal combustion engine has efficiency coefficient about 0.3 and a design can be done for a special flight light (without usual engine, transmission, wheels, etc.) car with small aerodynamic drag and AB engine has 100% efficiency and return the energy spent to lifting and acceleration of car. Moreover the AB car can fly in a straight line. If planet without atmosphere has natural or artifical magnetic field you can have a free flight to any planet place.

No problem to organize the air traffic for large numbers of the flying cars or flying people as is done for current aircraft and spacecraft. For example, in diapason of altitude 100-200 m flying cars move from West to East, in diapason 200 -300 m they move from North to South, in next diapason - from East to West and so on.

**3. Lavitation Aircraft with AB engine.** Design the aircraft having a flight weight 100 tons.
1. Take $m_p = 10^5$ kg, $B = 120$ T, $\gamma_s = 2300$ kg/m$^3$, $\gamma_w = 8000$ kg/m$^3$.
2. Computation: $w_v = 5.7 \times 10^9$ J/m$^3$, $P = 570$ kg/mm$^2$, $w_m = w_v/\gamma_s = 2.48 \times 10^6$ J/kg .
3. For $V = 250$ m/s = 900 km/h, $H = 30,000$ m, $L = 10,000$ km = $10^7$ m, $E_{in} = 3.6 \times 10^5$ W, $C_x = 0.06$, $S = 9.07$ m$^2$, $h = 14.5$ m (cylindrical tube is part of fuselage), we receive $D \approx 300$ N, $m_s = 2940$ kg.
4. Minimal cylinder radius and currency (for lift force of one cylinder $F = 10^6$ N): $R_m = 1.69$ m, $i = 10^9$ A.
5. Thickness of stuff: $\delta = 6.44 \times 10^{-3} \approx 6.44$ mm.
6. Cross-section and mass of wire (one tube) for currency density $j = 10^5$ A/mm$^2$: $s = 0.01$ m$^2$, $m_w = 112$ kg.
7. Total mass of levitation engine: $m = 2940 + 122 \approx 3062$ kg. Together with cooling system the mass of AB engine will be about 3.3 tons. The same way may be computed hypersonic or space aircraft. They can operate at very high altitude and have a small drag. That means they will spend very little energy per flight. If the aircraft will fly across space, the spent energy will be very small. We can also design the levitation space submarine with AB engine.

**4. Levitation (stasionary) Satellite with AB engine.** Compute AB engine for self-launch levitation stationary communication satellite located at an altitude of 40 km. This satellite can service a region within a radius of 700 km.
1. Take the data: magnetic intensity $B = 60$ T, $\gamma_s = 1800$ kg/m$^3$, $\gamma_w = 8000$ kg/m$^3$, form is cylinders, $h = 0.2$ m, $B_n = 3.4 \times 10^{-5}$ T, useful mass 80 kg.
2. Computation: $w_v = 1.4 \times 10^9$ J/m$^3$, $P = 140$ kg/mm$^2$, $w_m = w_v/\gamma_s = 7.8 \times 10^5$ J/kg .
3. For $V = 0$, $H = 40,000$ m, $L = 0$, $E_{in} = 0$, $k = 2$. we receive $m_s = 168.4$ kg for one cylinder.
4. Take the total mass of apparatus 280 kg. Minimal cylinder radius and currency (for lift force one cylinder $F = 2600$ N): $R_m = 0.63$ m, $i = 1.91 \times 10^8$ A.
5. Thickness of stuff: $\delta = 0,12$ m .
6. Cross-section and mass of wire (one tube) for currency density $j = 10^5$ A/mm$^2$: $s = 1.9 \times 10^{-3}$ m, $m_w = 1$ kg.
7. Mass of levitation engine : $m = 1.1 \times (168,4+1) \approx 184.1$ kg. Total mass of satellite is 184.1+80=264.1 $\approx$ 280 kg.

**6. Space Launch system.** Compute the AB space launcher for launching the spaceships of mass 20 tons to Moon
and Mars.
1. Take payload $m_p = 25$ tons $=25 \times 10^3$ kg, $B = 140$ T, $B_n = 3.4 \times 10^{-5}$ T, $\gamma_s = 2300$ kg/m$^3$, $\gamma_w = 8000$ kg/m$^3$, final speed $V = 11$ km/s, final altitude is $H = 200$ km, number of stages $N = 15$.
2. Compute: $w_v = 7.8 \times 10^9$ J/m$^3$, $P = 780$ kg/mm$^2$, $w_m = w_v/\gamma_s = 3.4 \times 10^6$ J/kg , $v_m = 2600$ m/s, $H_m = 340$ km.
3. Compute by Eq. (11) the mass of the last (N) stage: $a = 1.1724$, $m(N) = 272$ tons.
4. Compute the minimal radius and minimal currency of last stagy for length $h = 20$ m, $F = 2.72 \times 10^6$ N:
   $R_m = 2.86$ m, $i = 2 \times 10^9$ A, $m_s(N) = 40 \times 10^3$ kg.
5. Thickness of stuff: $\delta = 14 \times 10^{-3} \approx 14$ mm.
6. Cross-section and mass of wire for currency density $j = 10^5$ A/mm$^2$: $s = 0.02$ m$^2$, $m_w = 6400$ kg.
7. Total mass of the last stage without upper stages is 40 tons, total 272 tons.

All stages can be re-used for thousands of launchs. They can await the spaceship in flight and maintain levitation positions. When the spaceship returns and its mass is same to the launch mass, then all stages brake the spaceship and restore its energy. After landing they readied for the next free launch. Note, all stages are the thickness tubes which are inserted one into other.

**7. Artificial Magnetic field**. Let us take the supercoductive closed-loop cable-ring having radius $R$ = 10 km inside a big city. The efficiency radius of this ring is about 20 km or the diameter of artificial magnetic field is about 40 km. That is enough for any big city. If the cross-section area of the cable is 0.15 m$^2$ and currency density is $10^{11}$ A/m$^2$, the magnetic intensity $B = \mu_0 i/2R$ is about 1.9 T. Take for levitation devices the aluminum wire having $\rho = 2.8 \times 10^{-8}$ $\Omega$·m and $\gamma$ = 2800 kg/m$^3$, $j = 0.35 \times 10^6$ A/m$^2$, $B = 1.8$. From ratios (15) we receive

$$F_{kg}/m = 11.5 \text{ or } 115 \text{ N/kg}, \quad E/m = 2.45 \text{ W/kg}.$$

If mass of flying man is 80 kg and together with levitation device that is 100 kg, the mass of wire will be $m_w = 100/11.5 = 8.7$ kg and the heat expenses of energy in levitation is 245 W or $8.8 \times 10^5$ J in hours. The good rechargable battery has storage of an energy about 70 Wh/kg = $2.5 \times 10^5$ J/kg. That means the 8 kg battery has energy $E = 20 \times 10^5$ J. That is enough for two hours of flight. If it has a speed $V = 15$ m/s = 54 km/h, our range equals more 100 km. The air drag for this speed requires about $1.4 \times 10^5$ J. The full weight of levitation device is about 17 - 18 kg.

The levitation car may be computed similarly, but it needs a small conventional engine for support of car battery. If we are not limited to strong expenditures of energy, the current density may be increases from 0.35 A/mm$^2$ up to 10 A/mm$^2$. That decreases the mass of wire by 30 times, but increases the heat loss in wire by two orders.

The constant artificial magnetic field does not need in support energy. That also may be used as storage electric energy for big city. If we use the supercondutivity devices, the artificial magnetic field may be small.

If magnitic field is made variable and one direction of magnetic intensity (see Fig. 8b), this magnetic field will be able to pump the energy in the levitation engines and any special electric receivers having closed loop coil. We will not need a complex electric grid, which is in any big city. We can significantly improve the parameters of city ring and flight non-superconductivity devices if we use well ferro-magnetic matter. If we use permanent magnits for ground and apparatus, the human, vehicles can flight withotut spenging a large energy.

The initial data in all our projects are not optimal. Our aim - it shows that AB engine may be designed by current technology

## Discussion

The offered AB engines may be made by existing technology. We have a superconductivity material (see Table 1), the strong artificial fibers and whiskers (Table 2), the light cooling system (Table 3) for the Earth's atmosphere, and the radiation screens for outer space. The Earth has enough magnetic field, the Sun and many planets and their satellites (as Phobos orbiting Mars) has also magnetic field. The magnetic stars has powerful magnetic field and White Dwarfs have huge magnetic field (up $B = 80,000$ T). No problem to create the artificial magnetic field on asteroids and planet satellites (for example, to create local artificial magnetic field on the Moon). We have a very good perspective in improving our devices because—especially during the last 30 years—the critical temperature of the supercoductive material increases from 4 K to 186 K and no theoretical limit for further increase. Moreover, Russian scientists received the thin layers which have electric resistance at room temperature in million times less then the conventinal conductors. We have nanotubes whith will create the jump in AB engines, when their production will be cheaper. The current superconductive solenoids have the magnetic field $B \approx 20$ T.

AB engines can instigate a revolution in air, ground, sea and space transport. They allow individuals to fly as birds, almost energy-free (without loss of total energy or small expenses of energy) flight with hypersonic and space speed to any point of Earth and to other planets. The interstellar probes can use the magnetic fields of satellites and planet for braking and acceleration.

The AB engines solve the environment problem because they do not emit or evolve any polluting gases. They are useful in any solution for the oil-dependence problem because they use electricity and spend the energy for flight and other vehicles (cars) many times less than conventional internal combustion engine. In difference of a ground car, the levitation car flights in straight line to object.

The AB engines create the revolution in communication by the low altitude stationary suspended satellites, in energy industry, and especially in military aviation. They are very useful in lighting of

Earth by additional heat and light Sun radiation because, in difference from conventional mobile space mirrors, they can be suspended over given place (city) and service this place.

It is interesting, the toroidal AB engine is very comfortable for flying discs (UFO!) and have same property with UFOs. That can levitate and move in any direction with high acceleration without turning of vehicle, that does not excrete any gas, jet, that does not produce a noise.

Note, physicists have discussed for a long time the possible changing of weight the superconductivity magnet. Some of them are getting the changing and announce it as revolutionary discovery; others are repeating the test and getting negative results. The reason may be in different position their magnets and screening of part superconductivity coil about direction of the Earth's magnetic field.

## Conclusion

We must research and develop these ideas as soon as possible. They may to accelerate the technical progress and improve our life. There are no known scientific obstacles in the development and design of the AB engines, levitation vehicles, high speed aircraft, space launches, low aititude stationary communication satellites, cheap space trip to Moon and Mars and so on.

## Acknowledgement

The author wishes to acknowledge R.B. Cathcart for correcting the author's English and useful advice.